\documentclass[]{spie}  

\usepackage{amsmath,amsfonts,amssymb}
\usepackage{graphicx}
\usepackage[colorlinks=true, allcolors=blue]{hyperref}
\usepackage{xcolor}

\title{Long-term performance of SiPMs in space environment measured by GRBAlpha, GRBBeta, and VZLUSAT-2 CubeSats}

\author[a]{Jakub \v{R}\'{\i}pa}
\author[a]{Filip Münz}
\author[a]{Marianna Daf\v{c}\'{\i}kov\'a}
\author[b]{Andr\'as P\'al}
\author[a]{Norbert Werner}
\author[c]{Masanori Ohno}
\author[b]{L\'aszl\'o M\'esz\'aros}
\author[b]{Bal\'azs Cs\'ak}
\author[a]{Pavel Kosík}

\author[a]{Michaela \v{D}ur\'{\i}škov\'a}
\author[a]{Lea Szakszonov\'a}
\author[a]{Martin Kol\'a\v{r}}
\author[a]{Tom\'a\v{s} V\'{\i}tek}
\author[a]{Nikola Hus\'arikov\'a}
\author[a]{Filip Hroch}
\author[a]{Michal Pazderka}

\author[c]{Yasushi Fukazawa}
\author[c]{Hiromitsu Takahashi}
\author[c]{Tsunefumi Mizuno}
\author[c]{Jean-Paul Breuer}
\author[c]{Masato Yokota}
\author[d]{Kazuhiro Nakazawa}
\author[e]{Hirokazu Odaka}
\author[f]{Yuto Ichinohe}

\author[g]{Tom\'a\v{s} Urbanec}
\author[g]{Ale\v{s} Povala\v{c}}
\author[g]{Miroslav Kasal}

\author[h]{Miroslav \v{S}melko}
\author[i]{Peter Han\'ak}

\author[j]{G\'abor Galg\'oczi}
\author[k]{Martin Topinka}

\author[m]{Hsiang-Kuang Chang}
\author[n]{Tsung-Che Liu}
\author[o]{Chih-Hsun Lin}
\author[p]{Chin-Ping Hu}
\author[q]{Che-Chih Tsao}
\author[m]{Kaustubha Sen}
\author[m]{Chih-En Wu}

\author[r]{Jakub Kapu\v{s}}
\author[r]{J\'an Hudec}
\author[r]{Marcel Frajt}
\author[r]{Maksim Rezenov}

\author[s]{Vladim\'{\i}r D\'aniel}
\author[s]{Petr Svoboda}
\author[s]{Juraj Dud\'a\v{s}}
\author[s]{Martin Sabol}

\author[t]{Ivo Ve\v{r}t\'at}

\author[u]{Robert Laszlo}
\author[u]{Martin Koleda}

\affil[a]{Department of Theoretical Physics and Astrophysics, Faculty of Science, Masaryk University,  Kotl\'a\v{r}sk\'a 267/2, Brno 611 37, Czech Republic}
\affil[b]{Konkoly Observatory, Research Centre for Astronomy and Earth Sciences, Budapest, Hungary}
\affil[c]{Department of Physics, Graduate School of Advanced Science and Engineering, Kagamiyama, Hiroshima University, Higashi-Hiroshima, Japan}
\affil[d]{Department of Physics, Nagoya University, Nagoya, Aichi, Japan}
\affil[e]{Department of Physics, The University of Tokyo, Bunkyo-ku, Tokyo, Japan}
\affil[f]{RIKEN Nishina Center for Accelerator-Based Science, Saitama, Japan}
\affil[g]{Department of Radio Electronics, Faculty of Electrical Engineering and Communication, Brno University of Technology, Brno, Czech Republic}
\affil[h]{EDIS vvd., Ko\v{s}ice, Slovakia}
\affil[i]{Faculty of Aeronautics, Technical University of Kosice, Ko\v{s}ice, Slovakia}
\affil[j]{Wigner Research Centre for Physics, Budapest, Hungary}
\affil[k]{Charles University, Faculty of Mathematics and Physics, Institute of Astronomy, Prague, Czech Republic}
\affil[m]{Institute of Astronomy, National Tsing Hua University, Hsinchu, Taiwan, Republic of China}
\affil[n]{Department of Applied Physics, National Pingtung University, Pingtung, Taiwan, Republic of China}
\affil[o]{Institute of Physics, Academia Sinica, Taipei, Taiwan, Republic of China}
\affil[p]{Department of Physics, National Changhua University of Education, Changhua City, Taiwan, Republic of China}
\affil[q]{Department of Power Mechanical Engineering, National Tsing Hua University, Hsinchu, Taiwan, Republic of China}
\affil[r]{Spacemanic CZ, s.r.o., Brno, Czech Republic}
\affil[s]{VZLU AEROSPACE, a. s., Prague, Czech Republic}
\affil[t]{University of West Bohemia, Department of Applied Electronics and Telecommunications, Plze\v{n}, Czech Republic}
\affil[u]{NEEDRONIX s.r.o., Bratislava, Slovakia}

\authorinfo{Further author information: (Send correspondence to J. \v{R}\'{\i}pa)\\
J. \v{R}\'{\i}pa: E-mail: ripa.jakub@gmail.com}

\begin{document} 
\maketitle

\begin{abstract}
In this work, we report the successful application of silicon photomultipliers (SiPMs) in gamma-ray burst (GRB) detectors used in CubeSats operating in the low Earth orbit (LEO) radiation environment. It is known that SiPMs are susceptible to radiation damage, leading to an increase in the dark count rate. This results in an increase in the low-energy threshold in detectors combining SiPMs and scintillators. Despite this drawback, they became popular in gamma-ray detectors on CubeSats due to their low operating voltage, small size and fast response. Therefore, it is important to characterise their long-term performance in the space environment.
Here, we describe the changes in the dark count rate and low-energy threshold of S13360-3050PE multi-pixel photon counters (MPPCs) by Hamamatsu Photonics K.K., using measurements from the \textit{GRBAlpha}, \textit{GRBBeta}, and \textit{VZLUSAT-2} CubeSats. In the case of \textit{GRBAlpha}, the measurement of SiPM performance in space lasted over 4 years.
\textit{GRBAlpha} was a 1U CubeSat launched on 2021/03/22 to a 550 km altitude polar orbit carrying a CsI(Tl) scintillator GRB detector employing eight MPPCs and sensitive in the range of ~30-900 keV. \textit{GRBAlpha} de-orbited on 2025/06/09. \textit{VZLUSAT-2} was a 3U CubeSat launched on 2022/01/13 to a 535 km altitude polar orbit and de-orbited on 2025/11/30. \textit{GRBBeta} was launched on 2024/07/09 to a 580 km altitude, 62° inclination orbit. Both \textit{VZLUSAT-2} and \textit{GRBBeta} carry detectors similar to the one on \textit{GRBAlpha}.
We have flight-proven the Hamamatsu MPPCs S13360-3050 PE and demonstrated that SiPMs, shielded by 2.5 mm of PbSb alloy, can be used in a LEO environment on a scientific mission lasting beyond 4 years. This shows the potential for SiPMs to be employed in future satellites.
\end{abstract}

\keywords{gamma-ray detectors; SiPM; MPPC; radiation damage; space environment; LEO; CubeSat}

\section{Introduction}
\label{sec:intro} 
Silicon photomultipliers (SiPMs)\cite{2011NIMPA.656...69V, 2019NIMPA.926...16A, 2019NIMPA.926....2P, 2019NIMPA.926...36K} have become widely used in recent years in gamma-ray detectors on CubeSats due to their low operating voltage, small size, and fast response. Nonetheless, it is known that SiPMs are susceptible to proton and neutron radiation\cite{2003NIMPA.512...30L, 2019NIMPA.926...69G, 2023ExA....55..343M, 2021NIMPA.98864798M}, leading to an increase in the dark count rate, which affects their performance in space-based GRB detectors by increasing their low-energy sensitivity threshold.

In low Earth orbit (LEO), satellites face various components of particle radiation\cite{2019AN....340..666R, 2021JATIS...7b8004G, 2021SPIE11444E..3PR}, including energetic protons in the inner Van Allen radiation belt. Satellites at LEO pass this inner radiation belt in the region called the South Atlantic Anomaly (SAA). The SiPMs' performance degradation in space has been observed by several instruments: \textit{STPSat-5}/SIRI-1\cite{2021NIMPA.98864798M}, \textit{GRID-02}\cite{2022NIMPA104467510Z}, \textit{GECAM-A/B/C}\cite{2023NIMPA105668586Z, 2026RAA....26i5006L}, \textit{ÑuSat-7}/LabOSat-01\cite{2024NIMPA106769711F}, \textit{GRBAlpha} and \textit{VZLUSAT-2}\cite{2022SPIE12181E..1KR, 2025NIMPA107670513R}. The \textit{GRBAlpha}, \textit{VZLUSAT-2}, and \textit{GRBBeta} CubeSats have operated at LEO for several years, providing a good opportunity to study SiPMs' long-term behaviour in the space environment as summarised in this work.

\section{CubeSat missions}
\label{sec:cubesats}
The \textit{GRBAlpha}, \textit{VZLUSAT-2}, and \textit{GRBBeta} CubeSats carry on board gamma-ray burst (GRB) detectors, each detector consists of $75\times75\times5$\,mm$^3$ CsI(Tl) scintillator with a maximal on-axis effective area of about $50$\,cm$^2$ and each scintillator is read out by two sets of 4 SiPMs (multipixel photon counters S13360-3050PE by Hamamatsu Photonics K.K.). The scintillator is wrapped in Enhanced Specular Reflector (ESR) film, light-blocking polyvinyl fluoride (PVF) DuPont Tedlar, and sealed in an aluminium case. SiPMs are protected from protons by a 2--2.5\,mm thick PbSb alloy shield. Electronics features shaping amplifiers (15\,$\mu$s pulse width) and a Field-Programmable Gate Array (FPGA) to produce light curves.

All three CubeSat missions have been very successful in detecting GRBs and other gamma-ray transients. Over more than 4 years of observations, they detected 391 gamma-ray events, of which 28 were detected jointly. For a summary, see work\cite{2026arXiv260116609R}.

\subsection{GRBAlpha}
\label{sec:grbalpha}

\textit{GRBAlpha}\footnote{\href{https://grbalpha.konkoly.hu}{https://grbalpha.konkoly.hu}} (COSPAR ID: 2021-022AD)\cite{2020SPIE11444E..4VP, 2023A&A...677A..40P, 2023A&A...677L...2R} was a 1U CubeSat and a precursor of the proposed  mission CAMELOT\cite{2022SPIE12181E..1LM, 2021RMxAC..53..180M, 2018SPIE10699E..2PW, 2018SPIE10699E..64O}.
It had on board one GRB detector with the sensitivity in the range of $\sim 30-1\,000$\,keV. It was launched to an altitude of 550\,km, with a $97.5^\circ$ inclination, and operated for more than 4 years from 2021/03 to 2025/06, when it naturally de-orbited.

\subsection{VZLUSAT-2}
\label{sec:vzlusat2}

\textit{VZLUSAT-2}\footnote{\href{https://www.vzlusat2.cz}{https://www.vzlusat2.cz}} (COSPAR ID: 2022-002DF)\cite{2020SPIE11530E..0ZD, 2022Univ....8..241G} was a 3U CubeSat developed by VZLU AEROSPACE, a.s. It had, among several other payloads, two GRB detectors on board. The satellite had an active AOCS system that enabled control and determination of its attitude. It was launched to an altitude of 535\,km, with a $97.5^\circ$ inclination, and operated for almost 4 years from 2022/01 to 2025/10, when it naturally de-orbited.

\subsection{GRBBeta}
\label{sec:grbbeta} 

\textit{GRBBeta}\footnote{\href{https://grbbeta.tuke.sk}{https://grbbeta.tuke.sk}} (COSPAR ID: 2024-128C)\cite{2026arXiv260116609R} is a 2U CubeSat which has one GRB detector on board and an active AOCS system. It was launched to an altitude of 580\,km, with a $62^\circ$ inclination on 2024/07, and it is still operating at the time of writing this article.

\section{In-orbit measurement of the impact of radiation damage}
\label{sec:measurements}

In order to study the in-orbit degradation of the low-energy threshold of the GRB detectors and the dark count rate of SiPMs on the three CubeSats we regularly collected background spectra in low-background regions outside the Van Allen radiation belts with the highest possible spectral resolution (256 spectral channels) with the nominal operating bias voltage of SiPMs and typically accumulated over 60\,s.

\begin{figure}[h!]
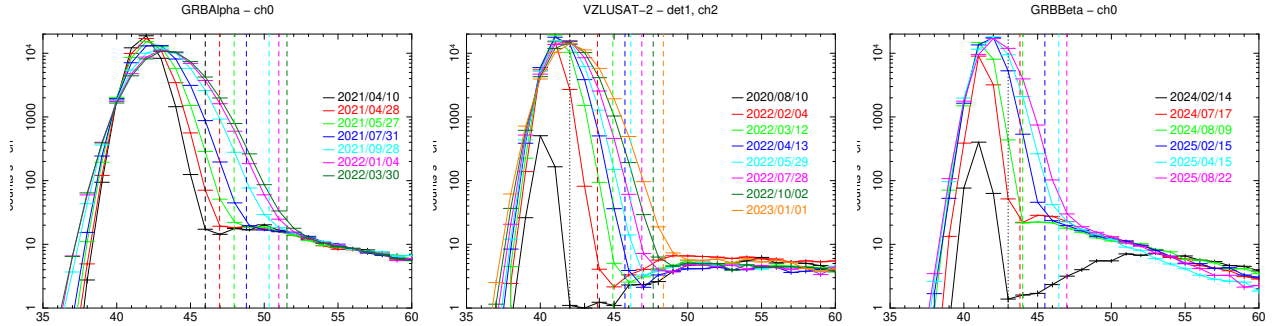

\begin{center}
\resizebox{0.32\textwidth}{!}{\includegraphics[trim={0 1.4cm 0 0.4cm},clip=true,angle=270]{images/GRBAlpha_spectra_channels3.eps}}\vspace*{2mm}
\resizebox{0.32\textwidth}{!}{\includegraphics[trim={0 1.4cm 0 0.4cm},clip=true,angle=270]{images/VZLUSAT2_spectra_channels2.eps}}\vspace*{2mm}
\resizebox{0.32\textwidth}{!}{\includegraphics[trim={0 1.4cm 0 0.4cm},clip=true,angle=270]{images/GRBBeta_spectra_channels.eps}}\vspace*{2mm}
\caption{Left: The noise peak evolution measured in-orbit by the GRB detectors on board of \textit{GRBAlpha} readout channel \emph{ch0} (left), \textit{VZLUSAT-2} GRB detector unit 1 readout channel \emph{ch2} (middle), and \textit{GRBBeta} readout channel \emph{ch0} (right) over the first year on orbit. For \textit{VZLUSAT-2} the spectrum shown in black (2020/08/10) and for \textit{GRBBeta} the spectrum shown in black (2024/02/14) were measured during the ground calibrations. The black dotted vertical lines mark the pre-launch thresholds, shown only for reference, as they were not calculated using the method that employs template spectra derived from post-launch measurements.}
\label{fig:noise_peak}
\end{center}
\end{figure}

The low-energy part of a measured background spectrum is dominated by the dark noise peak caused by thermal fluctuations of charge carriers. To determine the low-energy sensitivity threshold we use a method described in work \cite{2025NIMPA107670513R}. It employs the template background spectra derived from early post-launch measurements. Fig.~\ref{fig:noise_peak} presents the noise peak evolution observed in the background spectra of GRB detectors on board all three CubeSats over the first year in orbit.

\begin{figure}[h!]
\begin{center}
\resizebox{0.45\textwidth}{!}{\includegraphics{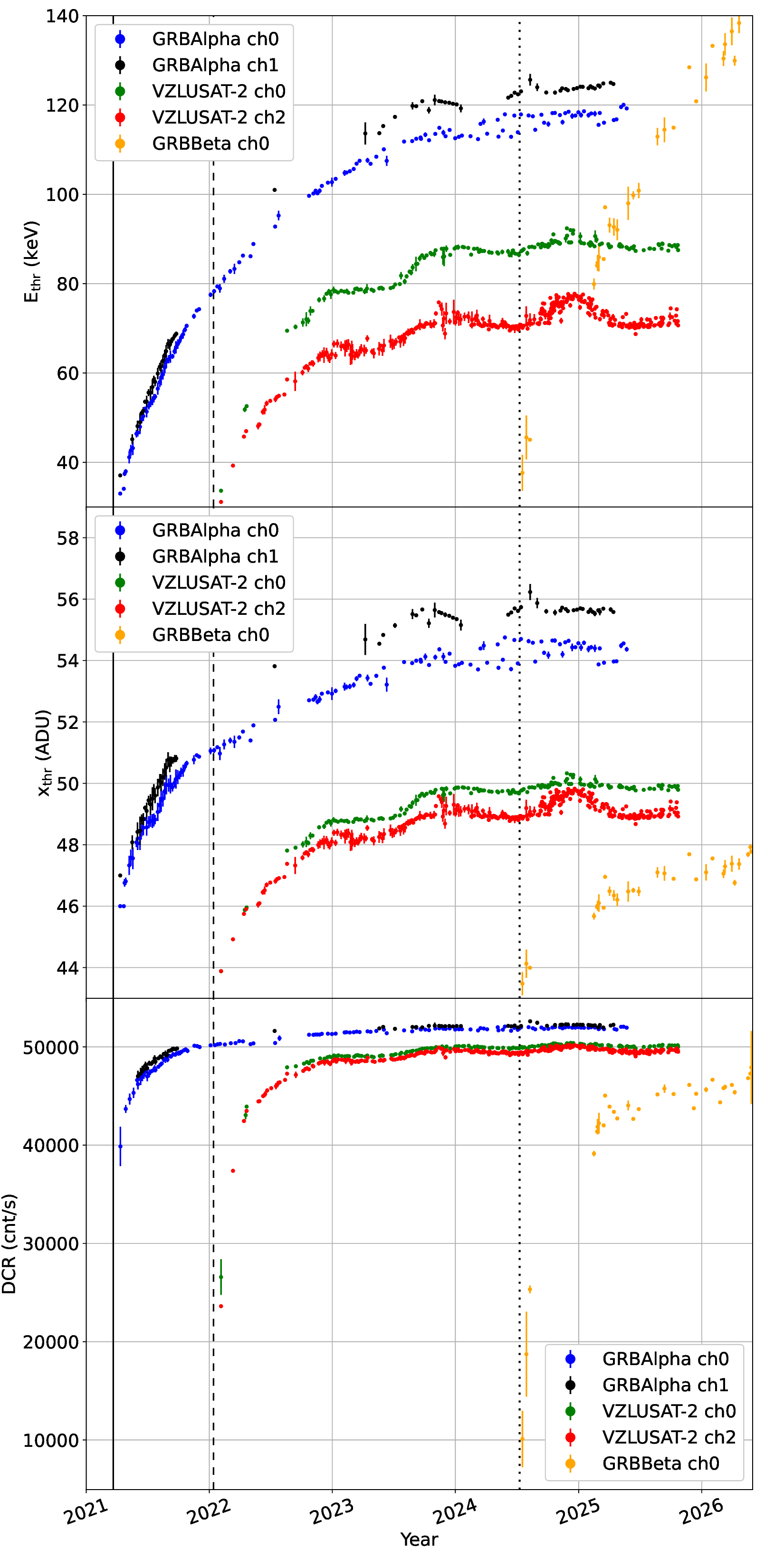}}\vspace*{2mm}
\caption{Top: Evolution of the low-energy threshold over time on orbit for the GRB detectors on the three CubeSats. Middle: Similar to the top panel, however here the threshold is plotted in terms of the directly measured spectral channel number (ADU). Bottom: Evolution of the dark count rate (DCR) integrated over the noise peak. The vertical solid, dashed, and dotted lines mark the launch dates of \textit{GRBAlpha}, \textit{VZLUSAT-2}, and \textit{GRBBeta}, respectively. The observed plateauing of the sensitivity threshold and DCR is due to lower proton flux at lower altitudes, resulting from natural orbital decay of the CubeSats. We caution that in-orbit gain calibration/correction for \textit{VZLUSAT-2} and \textit{GRBBeta} is not yet sufficiently accurate and hence the absolute values of the energy threshold are subject to uncertainty. See text for details.}
\label{fig:evolution}
\end{center}
\end{figure}

We noticed that the radiation damage of our SiPM-based detectors also manifests a substantial change in the detector's gain. Therefore, we had to recalibrate the detectors by analysing the shifts of the activation lines seen in the background spectra collected in the part of the orbit when a CubeSat passed SAA and moved northbound to regions with lower ambient radiation (for details see work\cite{2025NIMPA107670513R}). Fig.~\ref{fig:evolution} presents the evolution of the low-energy threshold over time on orbit for the GRB detectors on the three CubeSats. The figure shows the evolution of the threshold in terms of directly measured spectral channel number (ADU), in physical energy units derived from the ADU channel number using the gain calibration, and the evolution of the dark count rate (DCR), i.e. integrated number of counts over the noise peak.

In the case of \textit{GRBAlpha}, we had enough regularly recorded spectra with present activation lines, allowing us to reliably track the long-term gain change. However, note that we do not have enough measurements of the activation lines to track the long-term change of the gain of detectors on \textit{VZLUSAT-2} and we have to rely only on the pre-launch calibration from the ground. Therefore, we expect that the deviation of the low-energy threshold progress measured by \textit{VZLUSAT-2} from the measurement by \textit{GRBAlpha} is due to the absence of the correction for the in-orbit gain change. In the case of \textit{GRBBeta}, the in-orbit gain degradation correction is applied, but so far, the number of activation lines measurements is limited, and hence, the results of the energy threshold evolution are still preliminary and the derived low-energy sensitivity threshold converted from spectral channel number to keV should be taken with caution.

\section{Simulated in-orbit doses}
\label{sec:simulations}

To compare the measured low-energy degradation rate with the expected total ionising dose (TID) and total non-ionising dose or displacement damage (TNID) in SiPMs of the detectors, we run the following simulations (for details see work\cite{2025NIMPA107670513R}). We used the actual CubeSats' two-line element sets (TLE) obtained from the CelesTrak's satellite catalogue (SATCAT)\footnote{\href{www.celestrak.org/satcat/}{www.celestrak.org/satcat/}} to account for the progress of the natural satellite orbital decay.

Then we applied the \texttt{IRENE v1.57.004} software provided by the U.S. Air Force Research Laboratory
\footnote{\href{https://www.vdl.afrl.af.mil/programs/ae9ap9/}{https://www.vdl.afrl.af.mil/programs/ae9ap9/}} to calculate the omnidirectional differential particle fluence of geomagnetically trapped protons (AP-8 model). We choose the solar cycle activity minimum until Aug 2022, and from Sep 2022 onward, we choose the solar cycle activity maximum. Note that at LEO there is a lower flux of trapped protons during the solar cycle maximum compared to the solar cycle minimum \cite{1996GMS....97..119H}.

\begin{figure}[h!]
\begin{center}
\resizebox{!}{0.2\textheight}{\includegraphics{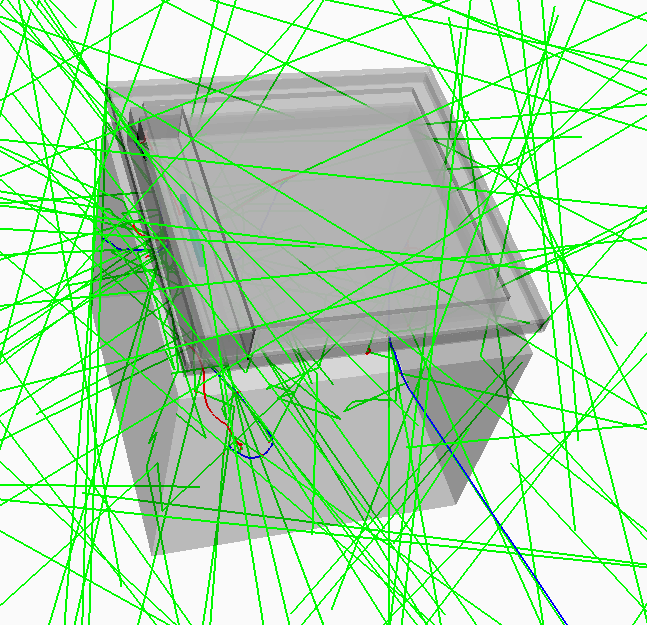}}\vspace*{2mm}
\resizebox{!}{0.2\textheight}{\includegraphics{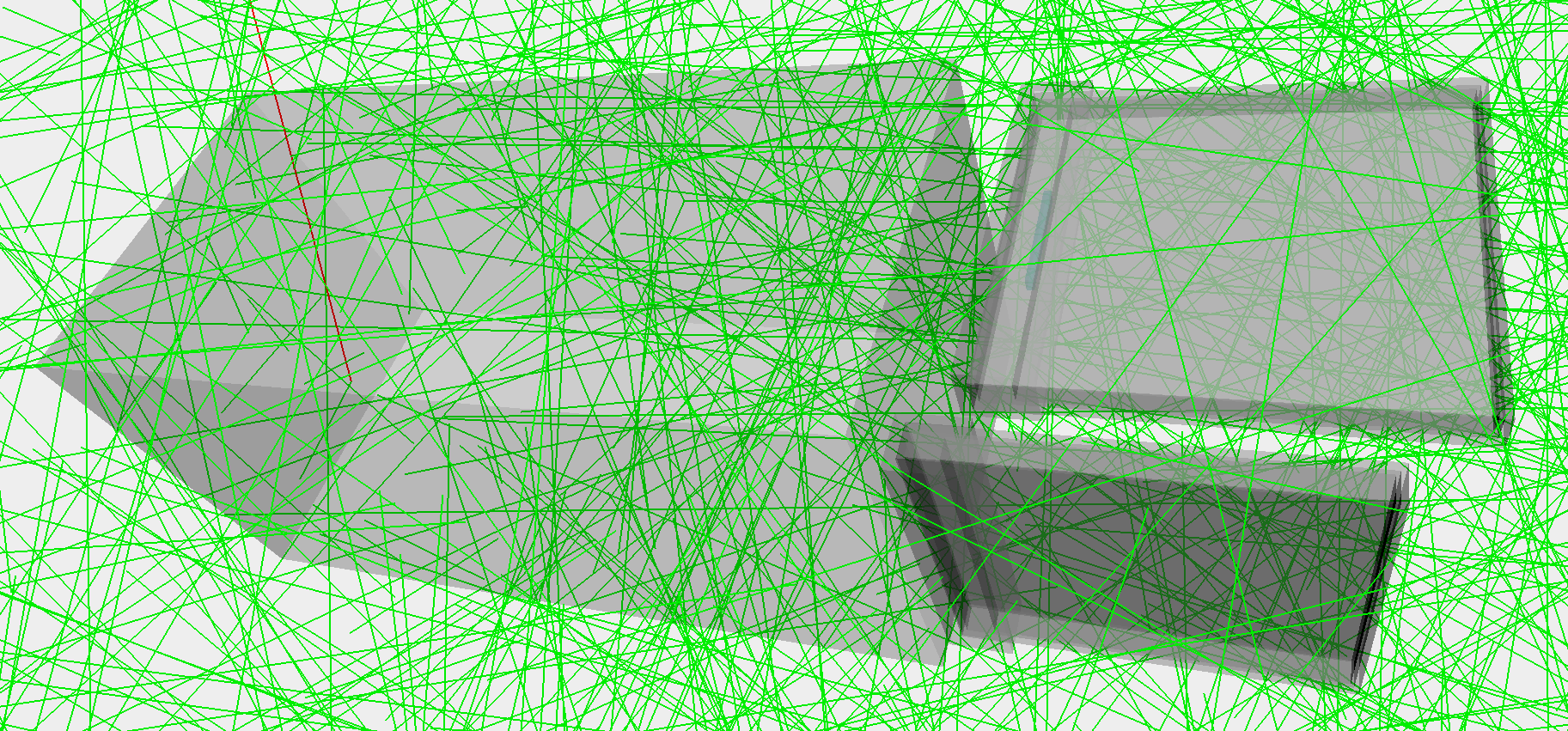}}\vspace*{2mm}
\caption{Example of the mass models of the CubeSats \textit{GRBBeta} (left) and \textit{VZLUSAT-2} (right) used to simulate TID and TNID in SiPMs with \texttt{GRAS} tool. Detectors were simulated in detail, and the remaining mass of each CubeSat was approximated as a box with corresponding mass. Tracks of the incident and secondary radiation (particles and gamma-rays) in the Geant4 simulation are visible, too.}
\label{fig:mass_models}
\end{center}
\end{figure}

Next, we simulated the TID and TNID deposited in Si of SiPMs using the ESA's Geant4 Radiation Analysis for Space \texttt{GRAS v06.00.01}
\footnote{\href{https://essr.esa.int/project/gras-geant4-radiation-analysis-for-space}{https://essr.esa.int/project/gras-geant4-radiation-analysis-for-space}}
software package\cite{2005ITNS...52.2294S} together with \texttt{Geant4 v.10.07.p01}\cite{2016NIMPA.835..186A} designed for analyses of radiation effects in materials, which works with complex 3D geometry models defined in Geometry Description Markup Language (GDML)\cite{2006ITNS...53.2892C}. For non-ionising energy-loss coefficients in Si, we chose the JPL/NRL/NASA (2003) values \cite{2003ITNS...50.1919M}. Fig.~\ref{fig:mass_models} shows the mass models of \textit{GRBBeta} and \textit{VZLUSAT-2} CubeSats used for these simulations.

\begin{figure}[h!]
\begin{center}
\resizebox{0.5\textwidth}{!}{\includegraphics{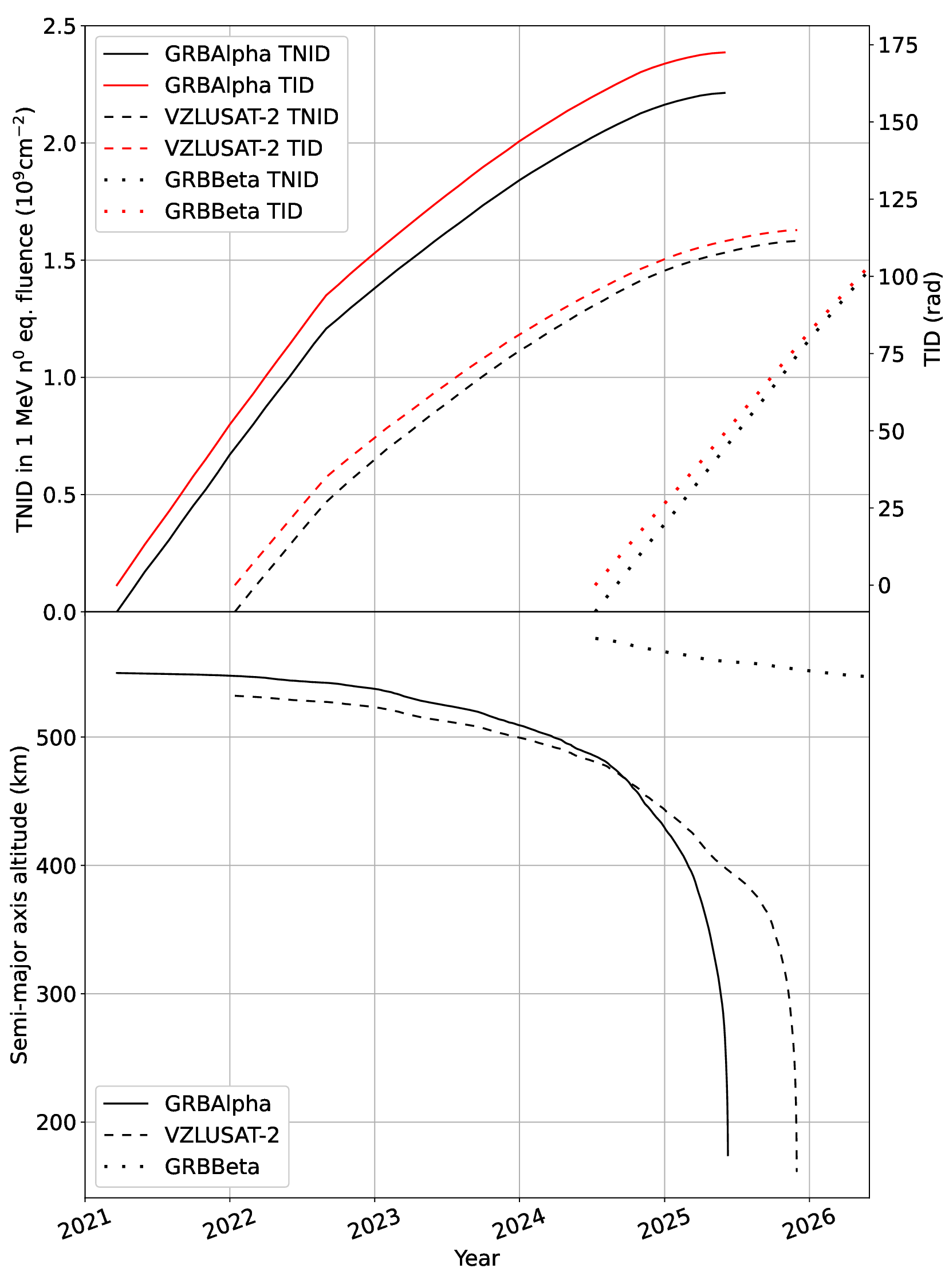}}\vspace*{2mm}
\caption{Top panel: Simulated total ionising dose and total non-ionising dose in SiPMs of GRB detectors of the three CubeSats. TNID is expressed in 1\,MeV neutron equivalent fluence. The AP-8 model of geomagnetically trapped protons was used in conjunction with actual satellite orbits and orbital attitude decay. Bottom panel: The progress of the semi-major axis altitude for the three CubeSats.}
\label{fig:doses_alt}
\end{center}
\end{figure}

Fig.~\ref{fig:doses_alt} presents the result of the simulated TID and TNID deposited in Si for the gamma-ray detectors on the three CubeSats together with the progress of the semi-major axis altitude.

\begin{figure}[h!]
\begin{center}
\resizebox{0.32\textwidth}{!}{\includegraphics{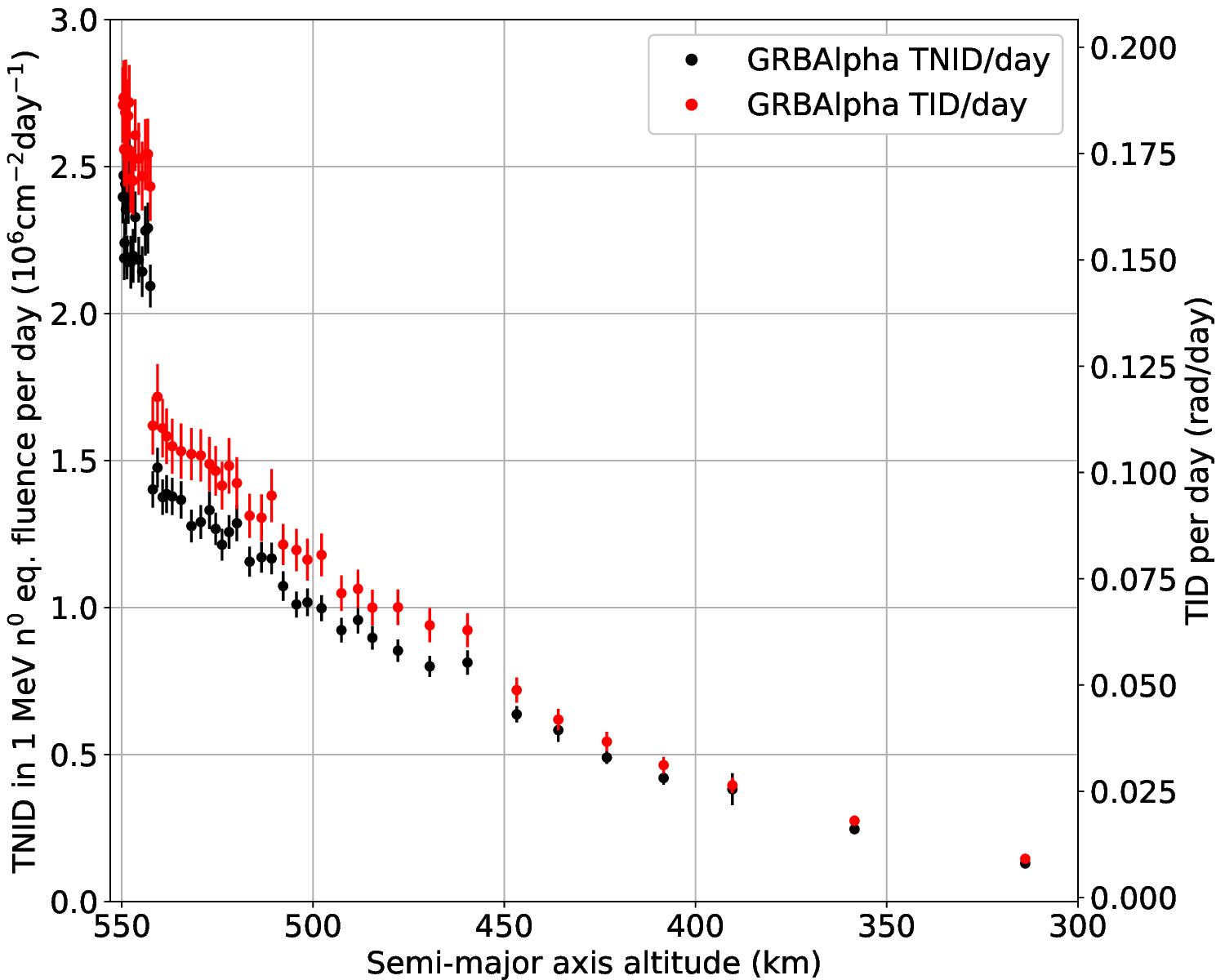}}\vspace*{2mm}
\resizebox{0.32\textwidth}{!}{\includegraphics{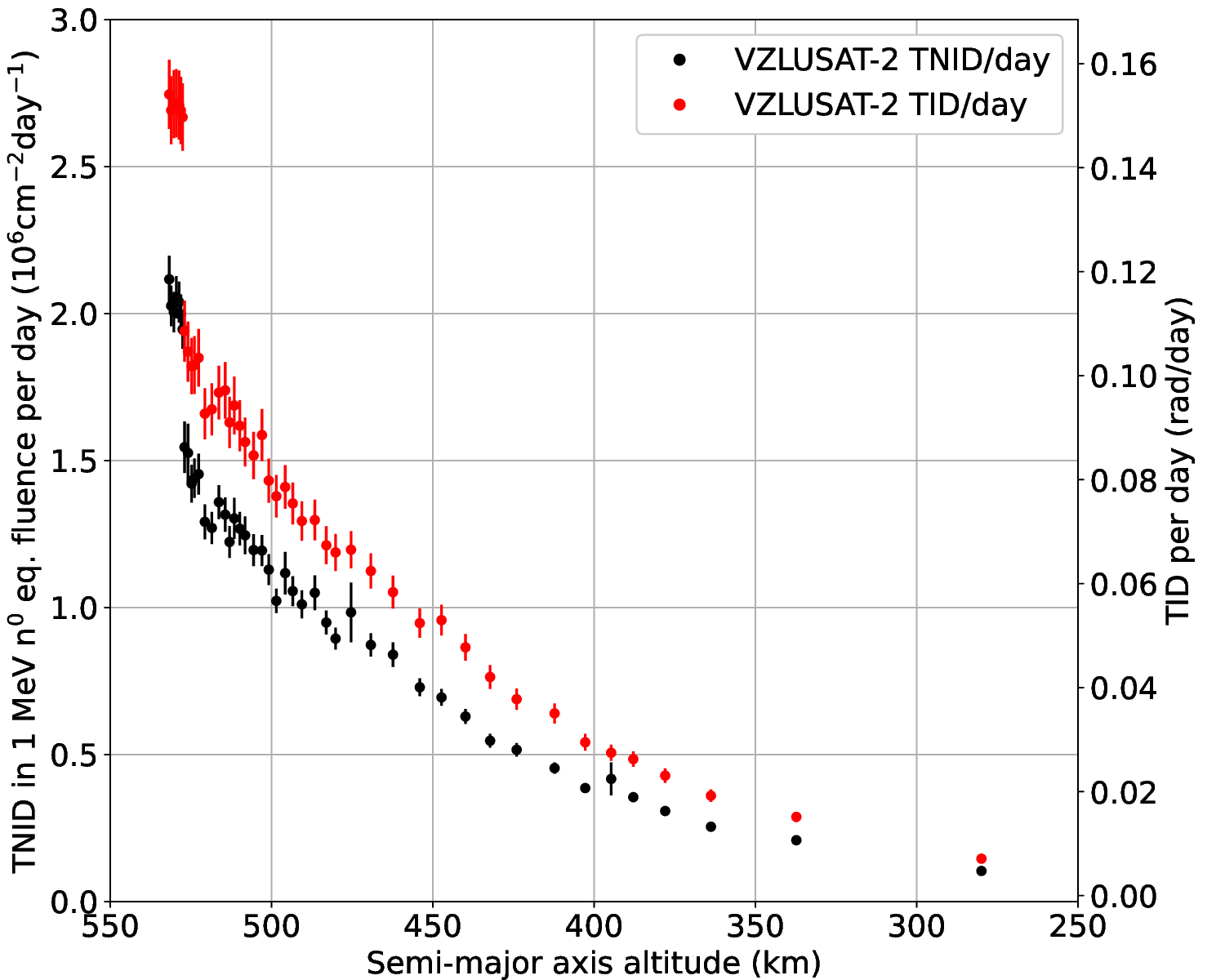}}\vspace*{2mm}
\resizebox{0.32\textwidth}{!}{\includegraphics{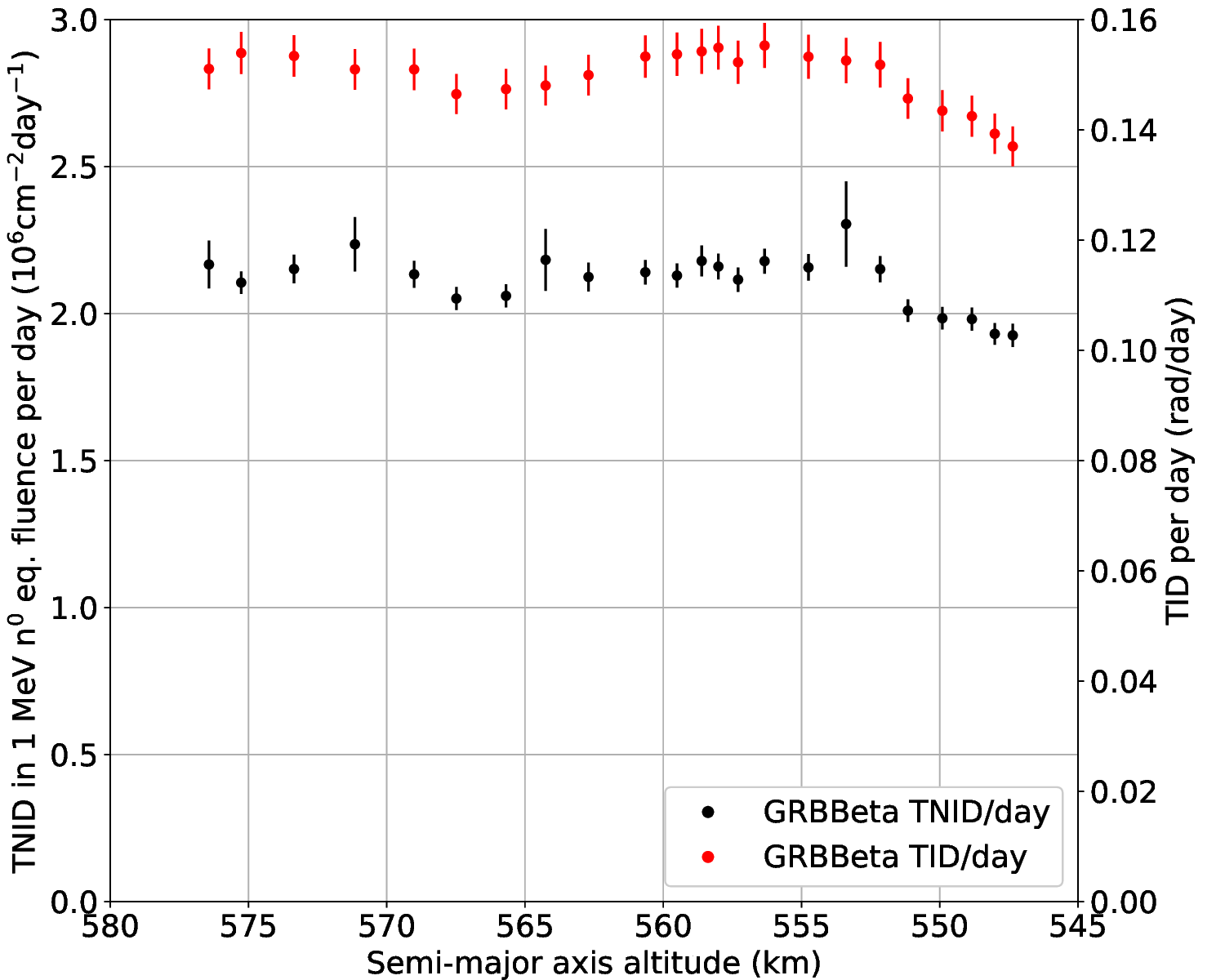}}\vspace*{2mm}
\caption{Total ionising dose per day and total non-ionising dose in 1\,MeV neutron equivalent fluence per day in Si simulated by the \texttt{GRAS} software using the AP-8 model of geomagnetically trapped protons and the actual orbital parameters of \textit{GRBAlpha} (left), \textit{VZLUSAT-2} (middle), and \textit{GRBBeta} (right).}
\label{fig:doses_daily}
\end{center}
\end{figure}

Fig.~\ref{fig:doses_daily} presents the effect of lower altitude on TID and TNID deposited per day in shielded Si as obtained from simulations including a transition from the solar cycle minimum to maximum.

\begin{figure}[h!]
\begin{center}
\resizebox{0.49\textwidth}{!}{\includegraphics{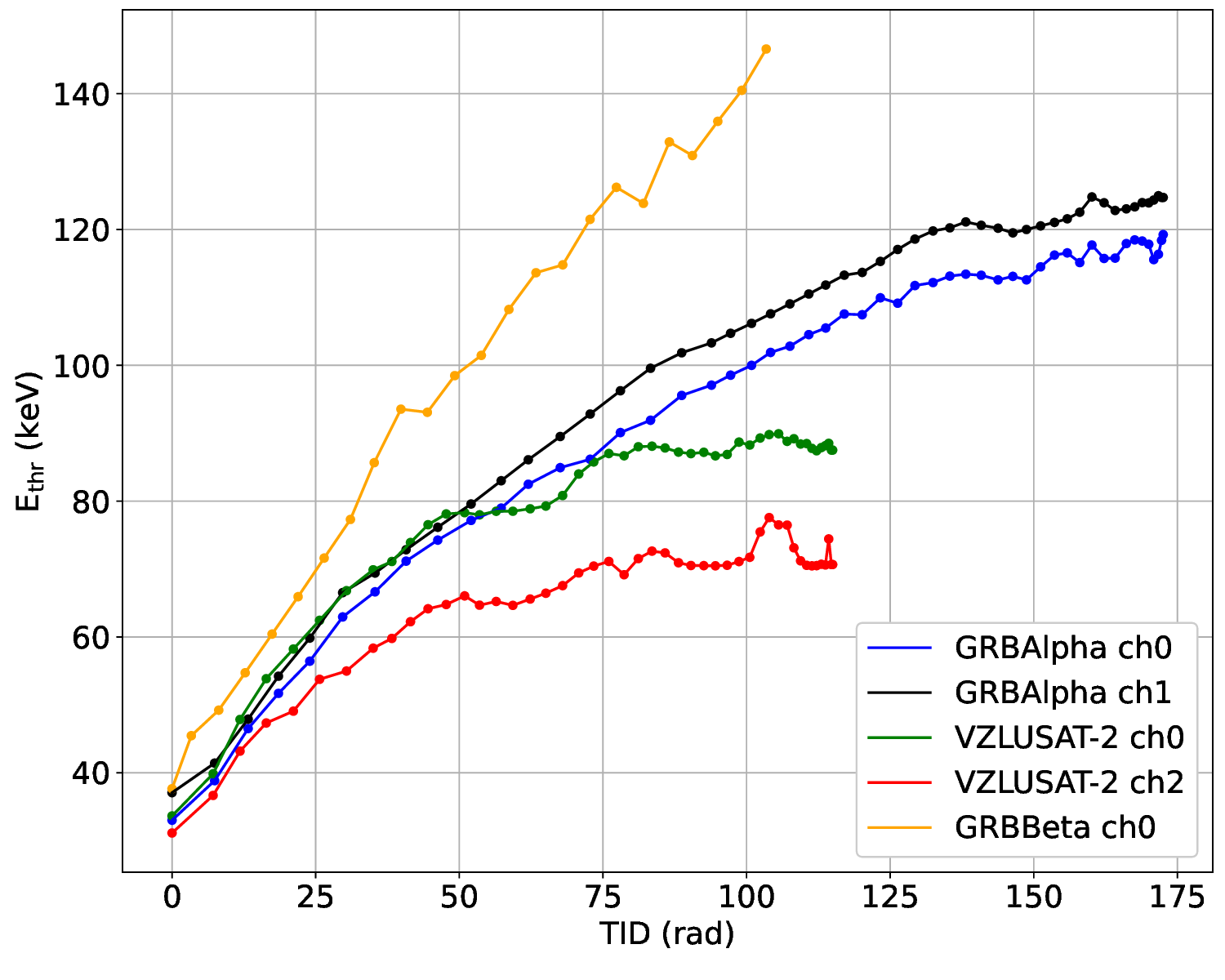}}\vspace*{2mm}
\resizebox{0.49\textwidth}{!}{\includegraphics{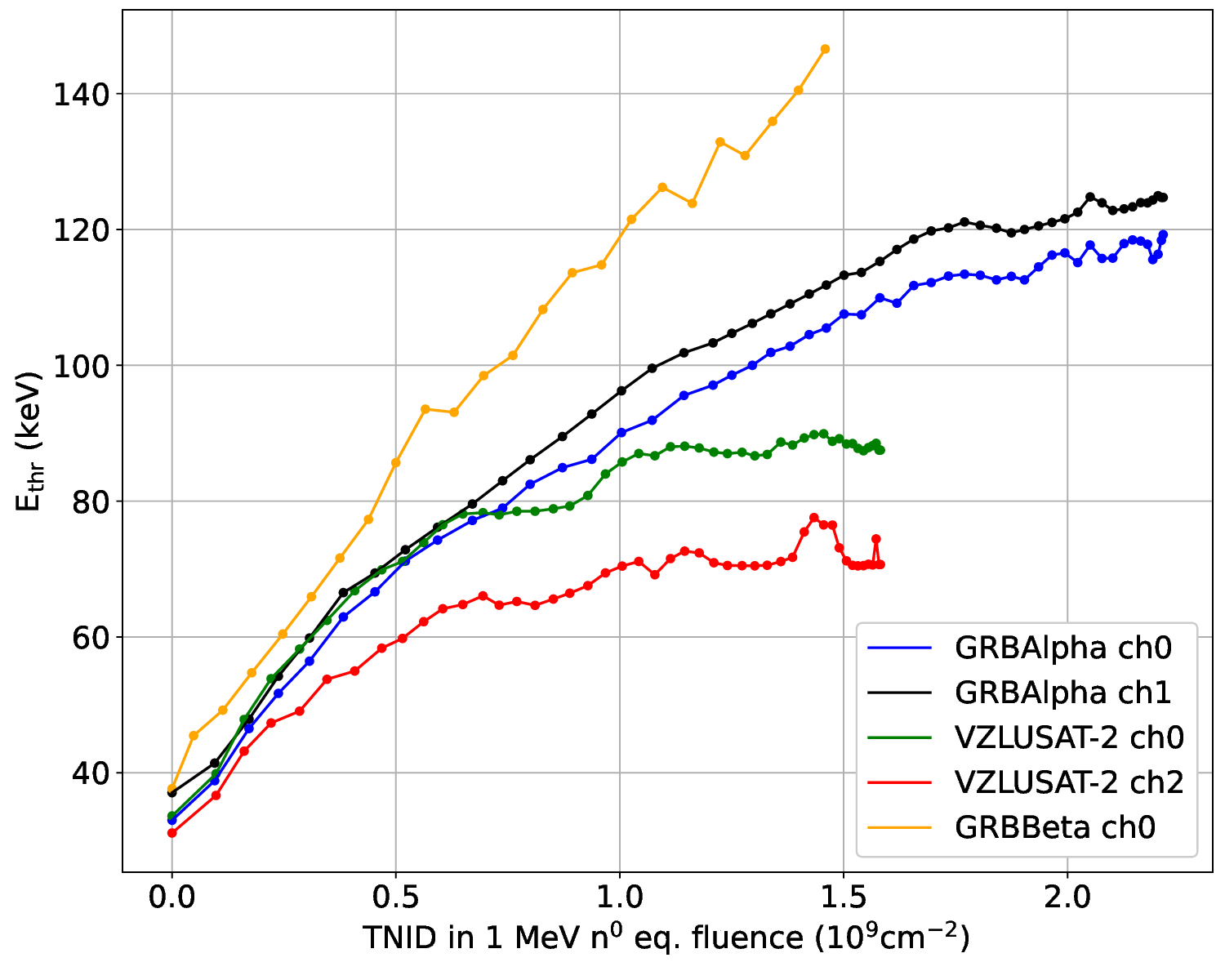}}\vspace*{2mm}
\caption{Dependence of the measured low-energy thresholds of SiPM-based gamma-ray detectors on \textit{GRBAlpha}, \textit{VZLUSAT-2} and \textit{GRBBeta} on the simulated TID (left) and TNID (right) derived from the mass model of the satellites, models of geomagnetically trapped protons, and the altitude decay progress over time.}
\label{fig:dose_thr}
\end{center}
\end{figure}

Fig.~\ref{fig:dose_thr} shows a comparison of the measured low-energy thresholds by \textit{GRBAlpha}, \textit{VZLUSAT-2} and \textit{GRBBeta} and the simulated in-orbit TID and TNID doses received by shielded SiPMs.

\section{Conclusions}
\label{sec:conclusions}
\begin{itemize}
    \itemsep-0.0em
    \item We have demonstrated that CubeSats can be used on missions at LEO lasting $>4$ years and routinely detect GRBs, solar flares and other gamma-ray transients.
    \item We have flight-proven compact GRB detectors based on CsI(Tl) scintillator read out by SiPMs by Hamamatsu Photonics K.K., MPPCs S13360-3050 PE, that can be readily implemented on CubeSat missions.
    \item We have demonstrated that SiPMs can be used in the LEO environment on a scientific mission lasting $>4$ years if sufficiently shielded. This demonstrates the potential of SiPMs for future high-energy astrophysics space missions.
\end{itemize}

\acknowledgments 
JR, FM, MD, and NW thank the support by the Czech Science Foundation (GAČR) project No. 24-11487J. MD is a Brno Ph.D. Talent Scholarship Holder---Funded by the Brno City Municipality. This work is also supported by the National Science and Technology Council (NSTC) of Taiwan under grants 113-2923-M-007-004-MY3.

\bibliography{main} 
\bibliographystyle{spiebib}

\end{document}